\documentclass[12pt]{article}
\usepackage{graphicx}
\def\hybrid{\topmargin 0pt      \oddsidemargin 0pt
        \headheight 0pt \headsep 0pt
        \voffset=-0.5cm
        \textwidth 6.25in       
        \textheight 9.5in       
        \marginparwidth 0.0in
        \parskip 5pt plus 1pt   \jot = 1.5ex}
\catcode`\@=11
\def\marginnote#1{}

\newcount\hour
\newcount\minute
\newtoks\amorpm
\hour=\time\divide\hour by60
\minute=\time{\multiply\hour by60 \global\advance\minute by-\hour}
\edef\standardtime{{\ifnum\hour<12 \global\amorpm={am}%
        \else\global\amorpm={pm}\advance\hour by-12 \fi
        \ifnum\hour=0 \hour=12 \fi
        \number\hour:\ifnum\minute<10 0\fi\number\minute\the\amorpm}}
\edef\militarytime{\number\hour:\ifnum\minute<10 0\fi\number\minute}

\def\draftlabel#1{{\@bsphack\if@filesw {\let\thepage\relax
   \xdef\@gtempa{\write\@auxout{\string
      \newlabel{#1}{{\@currentlabel}{\thepage}}}}}\@gtempa
   \if@nobreak \ifvmode\nobreak\fi\fi\fi\@esphack}
        \gdef\@eqnlabel{#1}}
\def\@eqnlabel{}
\def\@vacuum{}
\def\draftmarginnote#1{\marginpar{\raggedright\scriptsize\tt#1}}
\def\draftlabel#1{{\@bsphack\if@filesw {\let\thepage\relax
   \xdef\@gtempa{\write\@auxout{\string
      \newlabel{#1}{{\@currentlabel}{\thepage}}}}}\@gtempa
   \if@nobreak \ifvmode\nobreak\fi\fi\fi\@esphack}
        \gdef\@eqnlabel{#1}}
\def\@eqnlabel{}
\def\@vacuum{}
\def\draftmarginnote#1{\marginpar{\raggedright\scriptsize\tt#1}}

\def\draft{\oddsidemargin -.5truein
        \def\@oddfoot{\sl preliminary draft \hfil
        \rm\thepage\hfil\sl\today\quad\militarytime}
        \let\@evenfoot\@oddfoot \overfullrule 3pt
        \let\label=\draftlabel
        \let\marginnote=\draftmarginnote
   \def\@eqnnum{(\theequation)\rlap{\kern\marginparsep\tt\@eqnlabel}%
\global\let\@eqnlabel\@vacuum}  }

\def\numberbysection{\@addtoreset{equation}{section}
        \def\theequation{\thesection.\arabic{equation}}}

\def\underline#1{\relax\ifmmode\@@underline#1\else
        $\@@underline{\hbox{#1}}$\relax\fi}

\def\titlepage{\@restonecolfalse\if@twocolumn\@restonecoltrue\onecolumn
     \else \newpage \fi \thispagestyle{empty}\c@page\z@
        \def\thefootnote{\fnsymbol{footnote}} }

\def\endtitlepage{\if@restonecol\twocolumn \else  \fi
        \def\thefootnote{\arabic{footnote}}
        \setcounter{footnote}{0}}  
\relax

\hybrid

\newfont{\Bbb}{msbm10 scaled 1\@ptsize00}
\newcommand{\CC}{\mbox{\Bbb C}}

\newcommand{\TT}{\mbox{\Bbb T}}

\newcommand{\ZZ}{\mbox{\Bbb Z}}
\newfont{\Bbbb}{msbm7 scaled 1\@ptsize00}
\newcommand{\z}{\raise-1pt\hbox{$\mbox{\Bbbb Z}$}}

\def\beq{\begin{equation}}
\def\eeq{\end{equation}}
\def\p{\partial}

\begin{document}
\begin{titlepage}

\title{B\"acklund transformations
for difference Hirota equation and supersymmetric Bethe
ansatz\footnote{Based on the talk given at the Workshop ``Classical
and quantum integrable systems", Dubna, January 2007}}

\author{A.~Zabrodin
\thanks{Institute of Biochemical Physics,
4 Kosygina st., 119991, Moscow, Russia
and ITEP, 25 B.Cheremushkinskaya, 117259,
Moscow, Russia}}

\date{May 2007}
\maketitle

\begin{abstract}

We consider $GL(K|M)$-invariant integrable supersymmetric spin chains
with twisted boundary conditions
and elucidate the role of B\"acklund transformations in solving the
difference Hirota equation for eigenvalues of their transfer
matrices. The nested Bethe ansatz technique is shown to be
equivalent to a chain of successive B\"acklund transformations
``undressing" the original problem to a trivial one.

\end{abstract}

\vfill

\end{titlepage}

\section{Introduction}

For integrable models, the relationship between classical
and quantum systems is in no way exhausted by their correspondence
in the classical
limit. As is now well known, classical integrable equations
often appear in quantum integrable problems as exact relations
even for $\hbar \neq 0$.

One important example of this general phenomenon was investigated in
\cite{KLWZ,Z1,Z3}, where it was shown that the spectrum of commuting
transfer matrices (integrals of motion) in quantum integrable models
can be found in terms of discrete classical dynamics, also
integrable, defined in the space whose points label the commuting
transfer matrices. For integrable $GL(K)$-invariant spin chains
coordinates in this space are parameters specifying
finite-dimensional irreducible representations of the group $GL(K)$
and the spectral parameter. The classical dynamics in this space is
generated by functional relations for the transfer matrices
established in
\cite{BR1,KlumperPearce,Kuniba-1} for the ordinary bosonic case
and extended to the supersymmetric case in \cite{Tsuboi-1}.
Among them the most important is the bilinear functional equation
for the eigenvalues
of the transfer matrices ($T$-functions) which has the
form of the Hirota bilinear difference equation \cite{Hirota}. For
brevity, we call it the $TT$-relation. It is the starting point of
our approach.

The Hirota equation is probably the most famous equation in the
theory of classical integrable systems on the lattice. It provides a
universal integrable discretization of various soliton equations
and, at the same time, is a generating equation for their
hierarchies. It is involved in a large body of integrable problems,
classical and quantum. Like other soliton equations, the
Hirota equation admits (auto) B\"acklund transformations,
i.e., transformations that send any
solution to another solution of the same equation. They allow one
to construct a family of solutions that are connected with
a particularly simple one by a finite chain of such transformations.

The B\"acklund transformations play a central role in our method
serving as an alternative to the standard Bethe ansatz technique.
The nested Bethe ansatz solution of $GL(K)$-invariant spin chains
consists in successive increasing the rank of the group by applying
the Bethe ansatz repeatedly. In this way, one can descend from
$GL(K)$ to $GL(K\! -\! 1)$ until the problem gets trivialized at
$K=0$. At intermediate stages of this procedure, one introduces a
number of auxiliary ``$Q$-functions" (eigenvalues of Baxter's
$Q$-operators) connected with the $T$-functions by Baxter's
$TQ$-relations. Their zeros with respect to the spectral parameter
obey the system of Bethe equations. This purely quantum technique
has a remarkable classical interpretation \cite{KLWZ,Z1}: it is
equivalent to a chain of B\"acklund transformations for the Hirota
equation. The $TQ$-relations appear then as a constituent of
auxiliary linear problems for the Hirota equation. The rank of the
group becomes an additional discrete variable, with the dependence
on this variable being again described by the Hirota-like equation.
Since the solutions are polynomials in the spectral parameter, their
zeros obey equations of motion of a finite-dimensional
dynamical system in discrete time. The equations of motion are just
Bethe equations.

Recently, this approach was applied \cite{KSZ} to supersymmetric
spin chains constructed by means of $GL(K|M)$-invariant solutions to
the graded Yang-Baxter equation \cite{KulSk,Kulish}. In this case,
there are two rather then one additional discrete flows
corresponding to the bosonic and fermionic ranks, $K$ and $M$. Their
consistency leads to a non-trivial bilinear relation for eigenvalues
of Baxter's $Q$-operators (the $QQ$-relation). In the present paper
we extend these results to $GL(K|M)$-invariant spin chains with
twisted (quasi-periodic) boundary conditions. The twisting
parameters enter the solution as continuous parameters of the
B\"acklund transformations.

It is implied that the reader is familiar with the standard notions
and facts related to supergroups and their
representations \cite{Kac,Bars2,BMR}, as well
as with the quantum inverse scattering method constructions
\cite{FT,Faddeev,book}.

\section{The $TT$-relation}

Let us recall the construction of the family of commuting transfer
matrices in integrable spin chains. It is basically the same for
ordinary and supersymmetric models. We consider lattice models (spin
chains) with the symmetry supergroup $GL(K|M)$ constructed by means
of $GL(K|M)$-invariant $R$-matrices. Such $R$-matrices depend on the
spectral parameter $u\in \CC$ and act in the tensor product $V_0
\otimes V_1$ of two linear spaces, where irreducible representations
$\pi_0$ and $\pi_1$ of $GL(K|M)$ are defined. The space $V_0$ is
usually called auxiliary space and $V_1$ (local) quantum space. For
our purposes we need the case when $\pi_1$ is the vector
representation, i.e., $V_1= V= \CC^{K}\oplus \CC^{M}$ while $\pi_0$
is an arbitrary tensor representation of the supergroup. The
$R$-matrix reads
\beq\label{Rmat} R_{01}(u)=u+2\sum_{\alpha
\beta}(-1)^{p(\beta )} \pi_0(E_{\alpha \beta})\otimes e_{\beta
\alpha}\,.
\eeq
Here $p(\beta )$ is parity of the index $\beta$
($p(\beta ) =0$ or $1$), $e_{\alpha \beta}$ are matrices with the
entries $(e_{\alpha \beta})_{\alpha ' \beta '}= \delta_{\alpha
\alpha '} \delta_{\beta \beta '}$, and $\pi_0(E_{\alpha \beta})$ are
generators of the $gl(K|M)$ superalgebra in the representation
$\pi_0$. The first (scalar) term is to be understood as $u$
multiplied by the unity matrix $\pi_0 (I)\otimes I_{V_1}$, where
$I\in GL(K|M)$ is the unity element in the group and $I_{V_1}$ is
the identity operator in the space $V_1$. This $R$-matrix is the
$GL(K|M)$-invariant solution to the graded Yang-Baxter equation
\cite{KulSk,Kulish}. The supergroup invariance means that
\beq\label{inv}
\pi_0 (g)\otimes \pi_1 (g)\, R_{01}(u)= R_{01}(u) \,
\pi_0 (g)\otimes \pi_1 (g)
\eeq
for any $g\in GL(K|M)$.

In order to introduce generalized integrable spin chains on $N$ sites,
take $N$ copies of the space $V=V_1=V_2 =\ldots =
V_N$ (one for each site of the chain) and the corresponding $R$-matrices
$R_{0i}(u)$ acting in $V_0\otimes V_i$.
The Hilbert space of states of the model, ${\cal H}$,
is the tensor product of local quantum
spaces $V_i$ over all sites of the chain:
${\cal H}=\otimes_{i=1}^{N} V_i$.
We will call ${\cal H}$ the quantum space of the model.
The quantum monodromy
matrix is constructed as the product of the
$R$-matrices $R_{0i}(u)$ in the space $V_0$:
\beq\label{calT}
{\cal T}(u)=R_{01}(u-\xi_1)R_{02}(u-\xi_2)\ldots R_{0N}(u-\xi_N)\,.
\eeq
The quantities $\xi_i$ are input
data characterizing the (inhomogeneous) spin chain.
The supertrace of the quantum monodromy matrix taken
in the space $V_0$ gives a family of operators in the quantum space
depending on $u$ and $\pi_0$
(called transfer matrices),
which mutually commute for any values
of these parameters. A more general construction involves twisted
(quasi-periodic) boundary conditions defined by means of a diagonal matrix
\beq\label{gdiag}
g=\mbox{diag}\, (x_1, \ldots , x_K, \, y_1, \ldots , y_M)\in GL(K|M)
\eeq
(for simplicity, we identify elements of the supergroup $GL(K|M)$ with
matrices from its
vector representation). Then the commuting family of
transfer matrices is given by the formula
\beq\label{T1}
T^{(\pi_0)}(u;g)=\mbox{str}_{\pi_0} \left (\pi_0 (g){\cal T}(u)\right )\,.
\eeq
The graded Yang-Baxter equation combined with
the $GL(K|M)$ invariance implies that they commute for different
$u$ and $\pi_0$ (but not for different $g$!): $[T^{(\pi_0)}(u;g),\,
T^{(\pi_0')}(u';g)]=0$.

Below in this paper we will be especially interested in the case
when the representations $\pi_0$ are ``rectangular", i.e. correspond
to rectangular Young diagram. Given a rectangular diagram of
length $s$ and height $a$, let $\pi_{s}^{a}$ be the corresponding
representation. We define the quantum transfer matrices $T(a,s,u)$
for rectangular representations in the auxiliary space by the
formula
\beq\label{T2}
T(a,s,u)= \mbox{str}_{\pi_{s}^{a}} \Bigl
(\pi_{s}^{a} (g){\cal T}(u\! -\! s\! +\! a)\Bigr )\,.
\eeq
It
differs from (\ref{T1}) by the shift of $u$ which is convenient in
what follows. As a rule, we will not indicate the dependence on $g$
explicitly. One may formally extend this definition to zero values
of $a$ and $s$ which correspond to the trivial representation
$\pi_{0}^{a}=\pi_{s}^{0}=\pi_{\emptyset}$ ($\pi_{\emptyset}(g)=1$
for any $g\in GL(K|M)$). Taking into account that
$\pi_{\emptyset}(E_{\alpha \beta})=0$ for all generators of the
superalgebra, we conclude from formulas (\ref{Rmat}), (\ref{calT})
and (\ref{T2}) that
\beq\label{T3}
T(0,s,u)=\prod_{j=1}^{N}(u-s-\xi_j)\,, \quad
T(a,0,u)=\prod_{j=1}^{N}(u+a-\xi_j)
\eeq
where $\xi_j$ are the same
as in (\ref{calT}). So we see that $T(0,s,u)$ and $T(a,0,u)$ are
unity operators in the quantum space multiplied by scalar polynomial
functions. These functions are fixed input data of the problem. We put
\begin{equation}\label{F13}
\phi (u)=\prod_{i=1}^{N}(u-\xi_i)\,,
\end{equation}
then
$T(0,s,u)=\phi (u-s)$,
$T(a,0,u)=\phi (u+a)$.

The transfer matrices constructed above are linearly independent
but are connected by non-linear functional relations.
In particular,
the transfer matrices (\ref{T2})
for rectangular representations are known \cite{KlumperPearce,Kuniba-1}
to obey the $TT$-relation
\begin{equation}
\label{HIROTA}
  T(a,s,u \! +\! 1)T(a,s,u\! -\! 1)=
T(a,s \! +\! 1,u)T(a,s\! -\! 1,u) +T(a\! +\! 1,s,u)T(a\! -\! 1,s,u).
\end{equation}
This is the famous Hirota bilinear difference equation \cite{Hirota},
where $T$ plays the role of the $\tau$-function.
Since all the transfer matrices mutually commute, they can be
simultaneously diagonalized by a $u$-independent
similarity transformation,
and thus the same relation is valid for any of their
eigenvalues. Keeping this in mind, we will think of the transfer matrices
as scalar functions and call them $T$-functions.
Our strategy is to treat equation (\ref{HIROTA}) as the
basic equation of the quantum theory trying to derive the
results for the spectrum of quantum transfer matrices from it.
To do this, we need to specify the boundary conditions and
analytic properties of the solutions. The boundary conditions will be
discussed in section 4. The analytic properties in
$u$ are determined
by the type of the $R$-matrix (see a more detailed discussion
in \cite{Z3}).
For quantum spin chains with polynomial $R$-matrices (\ref{Rmat})
all the
$T$-functions
$T(a,s,u)$ are polynomials in $u$. Thus we are led to the
study of polynomial solutions to the Hirota equation.

One should note that the general solution to the Hirota
equation with the boundary $T$-functions at $a=0$ and $s=0$
as in (\ref{T3}) is given by the Bazhanov-Reshetikhin
determinant formula \cite{BR1}. In our normalization it
has the form
\beq\label{BRdet}
T(a,s,u)=H^{-1}(u\! -\! s, a)\det_{1\leq i,j\leq a}\,
T(1,\, s\! +\! i-j, \, u\! +\! a \! + \! 1\! - \! i\! -\! j)\,,
\eeq
where the function $H(u,a)$ is defined as follows:
$H(u,0)=1/\phi (u)$, $H(u,1)=1$,
$H(u,a)=\prod_{l=1}^{a-1}\phi (u+a-2l)$ at $a\geq 2$. It can be
expressed through the gamma-function for any $a$:
\beq\label{BRdet1}
H(u,a)=2^{(a-1)N}\prod_{i=1}^{N}
\frac{\Gamma \left (\frac{u+a-\xi_i}{2}\right )}{\Gamma
\left ( \frac{u-a-\xi_i}{2}+1\right )}\,.
\eeq
Given arbitrary functions $T(1,s,u)$,
formula (\ref{BRdet}) gives a solution to the Hirota equation.
However, in general this
solution does not obey the required analytical and boundary
conditions. In particular, the right hand side at $a\geq 2$
has apparent poles at zeros of the function $H(u-s, a)$.
Their cancelation is possible if the
polynomials $T(1,s,u)$ are chosen in a special way.

We conclude this section by a few words on normalization
of the solutions.
It is easy to check that the transformation
\begin{equation}\label{gaugef}
T(a,s,u)\longrightarrow f_0 (u \! +\! s \! +\!a)f_1 (u \! +\! s \!
-\!a) f_2 (u \! -\! s \! +\!a) f_3 (u \! -\! s \! -\!a) T(a,s,u),
\end{equation}
where $f_i$ are arbitrary functions, leaves the form of the Hirota
equation unchanged. One may choose certain normalization of the
solutions by fixing these functions in one or another way. In our
normalization, all the polynomials $T(a,s,u)$ have one and the same
degree $N$ equal to the number of sites in the spin chain. This
formally includes special cases when one or more zeros of some of
these polynomials are placed at infinity, then the degree is
actually less then $N$. Other ways of the normalization are
discussed in \cite{Z3,KSZ}.
In general position, the solutions
(\ref{T2}) are irreducible, i.e., $T(a,s,u)$ is not
divisible by any polynomial of the form $f_0 (u \! +\! s \! +\!a)f_1
(u \! +\! s \! -\!a) f_2 (u \! -\! s \! +\!a) f_3 (u \! -\! s \!
-\!a)$, where at least one of the polynomials $f_i (u)$ has degree
greater than $0$.

\section{The $N=0$ case: characters of the supergroup $GL(K|M)$}

Before proceeding further, it is instructive to consider
the case $N=0$ which appears to be
rather meaningful, though simple, and thus provides
useful analogies and motivations for dealing with more
complicated models. It also emerges as the $u\to \infty$
limit of models with $N>0$.

At $N=0$ there are no spin degrees of freedom and the $T$-functions
do not depend on $u$. As is seen from the definition, they coincide
with characters of the element $g\in GL(K|M)$  for rectangular
representations: $T(a,s,u)=\chi (a,s|g)$. The characters depend on
the parameters $x_i, y_j$ entering the matrix $g$ (\ref{gdiag}). We
assume that all  $x_i, y_j$ are distinct non-zero numbers. For
brevity, we will write $\chi (a,s)$ instead of $\chi (a,s|g)$ when
$g$ is fixed.

Let us introduce the rational function
\beq\label{ch1}
w(t)=
\frac{\prod_{m=1}^M (1-y_m t)}{\prod_{k=1}^K (1-x_k t)}\,,
\eeq
where $t$ is an auxiliary variable. As it follows from the character
formulas for supergroups \cite{Bars2}, $w(t)$ is the generating
function of the characters $\chi (1,s)$ while the inverse function
$w^{-1}(t)$ is the generating function of the characters $\chi
(a,1)$:
\begin{equation}
\label{ch2}
w(t)=\sum_{s=1}^{\infty} \chi (1,s)\, t^s\,,
\quad
w^{-1}(t)=\sum_{a=1}^{\infty} (-1)^a \chi (a,1)\, t^a\,.
\end{equation}
The other characters (super-analogs of Schur
functions) are expressed through $\chi (1,s)$ or
$\chi (a,1)$ by the Jacobi-Trudi determinant formulas:
\beq\label{ch3}
\chi (a,s)=\det_{1\leq i,j\leq a}\chi (1,\, s\! +\! i\! -\! j)=
\det_{1\leq i,j\leq s}\chi (a\! +\! i\! -\! j\,, \, 1)\,.
\eeq
They imply the bilinear relation for the characters
of rectangular representations:
\begin{equation}
\label{ch0}
\chi^2(a,s)=\chi(a+1,s)\chi(a-1,s)+\chi(a,s+1)\chi(a,s-1)
\end{equation}
which is the $u$-independent version of eq.~(\ref{HIROTA}). It is
known as the discrete KdV equation (see, e.g., \cite{DJM,Z2})
written in the bilinear form. Let us also mention the
integral representation of the characters
\beq\label{ch31}
\chi (a,s)=\frac{1}{(2\pi i)^a a!}\oint_{|t_1|=1}
\ldots \oint_{|t_a|=1}
\prod_{1\leq j<k\leq a}
|t_j -t_k |^2\,
\prod_{n=1}^{a}w(t_n) t_{n}^{-s-1}\, dt_n
\eeq
where it is assumed that all singularities of the
function $w(t)$ are outside the unit circle. This
$a$-fold integral coincides with the partition function
of the asymmetric unitary matrix model written through
the eigenvalues.

\begin{figure}[t]
   \centering
        \includegraphics[angle=-00,scale=0.5]{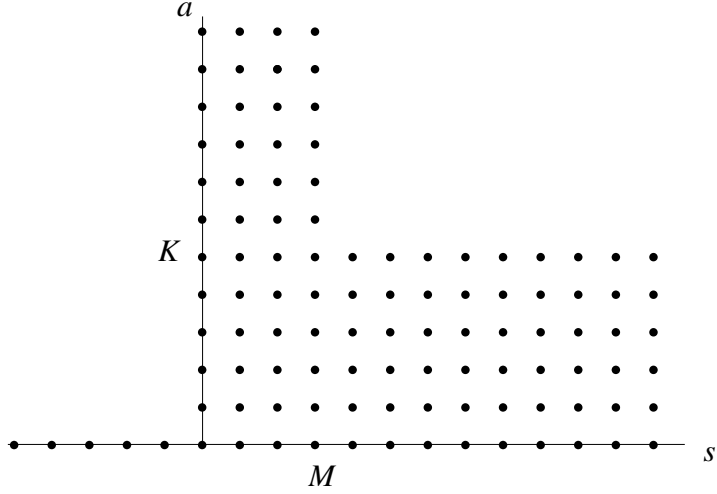}
        \caption{\it  The domain ${\sf H}(K|M)$
in the $(a,s)$-lattice at $K=5$, $M=3$.}
    \label{fig:HKM}
\end{figure}

One can formally extend the definition
of characters to negative values of $a,s$ by putting them
equal to zero. Because $\chi (0,s)=\chi (a,0)=1$ at
$a,s\geq 0$, this is consistent
with equation (\ref{ch0}) everywhere except at the point $a=s=0$.
To make the equation valid in the whole $(a,s)$ plane, one should
put either $\chi (0,n)=1$ or $\chi (n,0)=1$ for any $n\in \ZZ$, all
other $\chi (a,s)$ with negative $a$ or $s$ being zero. We choose
the former option (consistent with the integral
representation (\ref{ch31})).
Using representation (\ref{ch3}), it is not difficult to show
that if $a \geq K+1$ and $s\geq M+1$ then $\chi
(a,s)=0$. Summarizing, we have:
\beq \label{ch4}
\begin{array}{l}
  \chi (a,s)=0\quad  {\rm if:}\\ \\
\mbox{(i)} \;\; a<0
  \;\; {\rm or} \;\; \mbox{(ii)} \;\; a>0 \;\; {\rm and} \;\;
  s<0, \;\; {\rm or} \;\; \mbox{(iii)} \;\; a>K \;\; {\rm and}\;\;  s>M.
  \end{array}
\eeq
We see that the domain where $\chi (a,s)$ do not vanish
identically has the shape of a ``fat hook"  formed by
the union of the half-strips $a\geq 0, \, 0\leq s\leq M$ and
$s\geq 0, \, 0\leq a\leq K$
together with the horizontal ray $a=0, \, s<0$ from the origin to
minus infinity (see Fig.~\ref{fig:HKM}).
We denote this domain by ${\sf H}(K,M)$.
We call the boundaries at $a=0$ and $s=0, \, a\geq 0$
{\it exterior} and the boundaries inside the right upper
quadrant {\it interior} ones.

The characters on the interior boundaries can be found explicitly. A
simple calculation shows
that $\chi (K, M+n)$ and $\chi (K+n, M)$ at
$n\geq 0$ are given by the formulas
\beq\label{ch6}
\chi (K,M+n)=\left (\prod_{k=1}^{K}x_{k}^{n}\right )
\prod_{i=1}^{K}\prod_{j=1}^{M}(x_i - y_j ),
\eeq
\beq\label{ch6a}
\chi (K+n,M)=\left (\prod_{m=1}^{M}(-y_{m})^{n}\right )
\prod_{i=1}^{K}\prod_{j=1}^{M}(x_i - y_j )
\eeq
(in fact this is a particular case of a more general
factorization property, see \cite{Macdonald}, chapter I,
section 3, example 23).
Therefore, they
are connected by the relation
\begin{equation} \label{ch}
  \chi (K,M+n)=(-1)^{nM} (\mbox{sdet}\, g)^n \,
  \chi (K+n,M),\quad  n\ge 0,
\end{equation}
where
$$
\mbox{sdet}\, g=\frac{x_1 x_2\ldots x_K}{y_1 y_2\ldots y_M}\,.
$$

Using the determinant identities (Pl\"ucker relations) for minors of
a rectangular matrix built from the elementary characters
$\chi (1, s)$, one can prove the three-term bilinear relations
\beq\label{ch8}
\begin{array}{l}
\chi (a+1,s)\tilde \chi (a,s)-
\chi (a,s)\tilde \chi (a+1,s)=z \chi (a+1,
s-1)\tilde \chi (a,s+1)\,,
\\ \\
\chi (a,s+1)\tilde \chi (a,s)-\chi (a,s)\tilde \chi (a,s+1)
=z\chi (a+1, s)\tilde \chi (a-1, s+1)
\end{array}
\eeq
between the characters $\chi (a,s )$ with the generating
function $w(t)$ and the characters
$\tilde \chi (a,s)$ with the generating
function $\tilde w(t)=(1-zt)w(t)$.  Both $\chi$ and $\tilde \chi$
obey the discrete KdV equation (\ref{ch0}) and thus relations
(\ref{ch8}) generate the B\"acklund transformation for it.
Their meaning is to relate characters of
$GL(k|m)$-representations with different $k,m$.
Indeed, let $g_{k,m}\in GL (k|m)$ be the diagonal matrix
\beq\label{gkmdiag}
g_{k,m}=\mbox{diag}\, (x_1, \ldots , x_k , y_1, \ldots , y_m)
\eeq
obtained from
$g=g_{K,M}$ (\ref{gdiag}) by removing $K-k$ eigenvalues $x_{K},
x_{K-1}, \ldots , x_{k+1}$ and $M-m$ eigenvalues $y_{M}, y_{M-1},
\ldots , y_{m+1}$. We put
\beq\label{ch7}
\chi_{k,m}(a,s)=\chi (a,s |g_{k,m})
\eeq
and, in the same way as in (\ref{ch1}), (\ref{ch2}),
introduce the generating function of the characters
$\chi_{k,m}(1,s)$:
\beq\label{ch1a}
w_{k,m}(t)=
\frac{\prod_{i=1}^m (1-y_i t)}{\prod_{j=1}^k (1-x_j t)}\,.
\eeq
The obvious
recurrence relations
\beq\label{rec}
\begin{array}{l}
w_{k-1,m}(t)=(1-x_k t)\, w_{k,m}(t)\,,
\\ \\
w_{k, m+1}(t)=(1-y_{m+1}t)\, w_{k,m}(t)
\end{array}
\eeq
allow one to apply equations (\ref{ch8}), where one should put
$\chi (a,s)=\chi_{k,m}(a,s)$, $\tilde \chi (a,s)=
\chi_{k-1,m}(a,s)$ at $z=x_k$ or  $\chi (a,s)=\chi_{k,m-1}(a,s)$,
$\tilde \chi (a,s)=\chi_{k,m}(a,s)$ at $z=y_m$.

\section{The boundary conditions}

In general, the Hirota equation has many solutions of
very different natures. The most important additional ingredient which
selects the class of solutions relevant to quantum integrable
models is the boundary conditions in the variables $a,s$.
Qualitatively, these conditions are the same as those for
characters of rectangular representations of supergroups discussed
in the previous section, and can be derived by means of a similar
reasoning. Furthermore, the explicit formula (\ref{Rmat}) for the
$R$-matrix implies that the highest coefficient of the polynomial
$T$-function coincides with the
corresponding character: $T(a,s,u)=\chi (a,s)u^N
+O(u^{N-1})$ as $u\to \infty$.

The distinctive feature of the solutions of our interest is that they
are required to vanish identically in some parts of the
$(a,s)$-plane. The domain where they do not do so
depends on the symmetry algebra of the quantum model.
For the $GL(K|M)$-invariant supersymmetric spin chains
it is the domain ${\sf H}(K,M)$ introduced in the previous section
(Fig.~\ref{fig:HKM}). (For
the $GL(K)$-invariant spin chains it degenerates to the half-strip
$0\leq a\leq K$, $s\geq 0$ together with the two rays $s=0$, $a\geq
K$ and $a=0$, $s\leq 0$.)
Similarly to (\ref{ch4}), we can write:
\beq \label{BCOND1}
\begin{array}{l}
  T(a,s,u)=0\quad  {\rm if:} \\ \\
\mbox{(i)} \;\; a<0
  \;\; {\rm or} \;\; \mbox{(ii)} \;\; a>0 \;\; {\rm and} \;\;
  s<0, \;\; {\rm or} \;\; \mbox{(iii)} \;\; a>K \;\; {\rm and}\;\;  s>M.
  \end{array}
\eeq
As is easy to see, the shape of ${\sf H}(K,M)$ is consistent
with the Hirota equation in the whole $(a,s)$-plane. Although only
the points with non-negative $a,s$ have the direct physical
interpretation, it is important to consider the full domain (the
``fat hook" complemented by the ray) since otherwise the Hirota
equation would brake down at the corner point at the origin.

The boundary  values of $T$-functions have a rather special
factorized form fixed by consistency with the Hirota equation.
Indeed, on the boundary one of the two terms in the right hand
side of eq. (\ref{HIROTA}) vanishes resulting in constraints
on the boundary values.
For example,
at $a=0$ eq. (\ref{HIROTA}) converts into
$$
T(0,s,u \! +\! 1)T(0,s,u\! -\! 1) = T(0,s \! +\! 1,u)T(0,s\! -\!
1,u)
$$
which is a discrete version of the d'Alembert equation with the
general solution $T(0,s,u)=f_{+}(u+s) f_{-}(u-s)$ where $f_{\pm}$
are arbitrary functions. In a similar way,
$T(a,0,u)=\tilde f_{+}(u+a) \tilde f_{-}(u-a)$. Equations (\ref{T3})
show that in our normalization $f_{+}(u)=\tilde f_{-}(u)=1$ while
$f_{-}(u)=\tilde f_{+}(u)=\phi (u)$, where $\phi (u)$ is defined
in (\ref{F13}).
A similar factorization holds true
on the interior boundaries. However, in general both functions in the
product are non-trivial and are to be determined from the
$TT$-relation. The case of the ordinary group
$GL(K)=GL(K|0)$, when
the vertical interior boundary coincides with the exterior one,
is special in this respect. In this case all the boundary
values enter as fixed input data. In particular,
on the horizontal interior boundary we have
$T(K, s, u)=\phi (u+s+K)$.

On the interior boundaries, we impose the condition
\begin{equation} \label{IDENT}
  T(K,M+n,u)=(-1)^{nM} (\mbox{sdet}\, g)^n \,
  T(K+n,M,u),\quad  n\ge 0,
\end{equation}
which is consistent with the corresponding condition (\ref{ch})
for characters
of the supergroup.
For the periodic case (when $g$ is the unity matrix)
this condition was pointed out in
\cite{Tsuboi-1}.
Equality (\ref{IDENT}) means that the values of the $T$-functions at
the points of the interior boundaries equally spaced from the
corner point differ by a constant factor only. Note that this
condition trivially holds in the case $M=0$, where
these values are fixed from the
very beginning as input data. As it was already
mentioned, at any $K,M>0$ they are no longer
fixed but are to be determined together with the
$T$-functions inside the domain ${\sf H}(K|M)$.

Concluding this section, we note that the parameters
$x_k, \, y_m$ do not enter explicitly neither the Hirota
equation nor the boundary conditions for it. This means
that the Hirota equation with the fixed boundary
conditions on the exterior boundaries of the form
(\ref{T3}) has a continuous $(K+M)$-parametric family of
polynomial solutions.

\section{Auxiliary linear problems}

Like almost all known nonlinear integrable equations, the Hirota
equation is a compatibility condition for an
over-determined system of linear problems \cite{SS,KLWZ}. To introduce them,
it is convenient to pass to the ``chiral" variables
\begin{equation}\label{LP1}
\begin{array}{l}
p=\frac{1}{2}(u-s-a)\\ \\
q=\frac{1}{2}(u+s+a)\\ \\
r=\frac{1}{2}(-u-s+a)\,.
\end{array}
\end{equation}
The original variables $a,s,u$ will be referred to as ``laboratory" ones.
Here are the
formulas for the inverse transformation,
\begin{equation}\label{LP2}
a=q+r\,, \quad s=-p-r\,, \quad u=p+q
\end{equation}
and for the transformation of the vector fields:
\begin{equation}\label{LP3a}
\p_p =\p_u - \p_s\,, \quad \p_q =\p_u + \p_a\,, \quad \p_r =\p_a -
\p_s
\end{equation}

We set $\tau (p,q,r)= T(q+r,\,  -p-r, \, p+q)$ and introduce the
following linear problems for an auxiliary function $\psi = \psi
(p,q,r)$:
\begin{equation}\label{LP3}
\begin{array}{l}
\displaystyle{ \psi (r+1) + z \frac{\tau (p\! +\! 1, r\! +\!
1)\,\, \tau}{\tau (p+1)\tau (r+1)}
\, \psi = \psi (p+1)} \\ \\
\displaystyle{ \psi (r+1) - z \frac{\tau (q\! +\! 1, r\! +\!
1)\,\, \tau}{\tau (q+1)\tau (r+1)} \, \psi = \psi (q+1)}
\end{array}
\end{equation}
where $z$ is a parameter. In these formulas,
we indicate explicitly only those variables that are subject
to shifts. Using these equations, the function $\psi (p+1, q+1)$ can be
represented as a linear combination of $\psi (r)$, $\psi (r+1)$ and
$\psi (r+2)$ in two different ways. Compatibility of
the linear problems means that the
results are to be equal. Equating the two expressions we see that
the terms proportional to $z^2\psi (r)$ and $\psi (r+2)$
cancel automatically while the terms proportional to $z\psi (r+1)$
yield a non-trivial relation (provided $\psi (r+1)$ does not vanish)
$$
\frac{\tau (p\! +\! 1, r\! +\! 2)\,\, \tau (r+1)}{\tau
(p\! +\! 1, r\! +\! 1)\tau (r+2)} \, -\,
\frac{\tau (p\! +\! 1, q\! +\! 1, r\! +\! 1)\tau (p+1)}{\tau
(p\! +\! 1, q\! +\! 1)(p\! +\! 1, r\! +\! 1)}
$$
$$
=\,\,\, \frac{\tau (p\! +\! 1, q\! +\! 1, r\! +\! 1)\tau (q+1)}{\tau
(p\! +\! 1, q\! +\! 1)(q\! +\! 1, r\! +\! 1)}\, -\, \frac{\tau (q\!
+\! 1, r\! +\! 2)\,\, \tau (r+1)}{\tau (q\! +\! 1, r\! +\! 1)\tau
(r+2)}\,.
$$
This equality states that the function
$$
\frac{\tau (p+1)\tau (q+1, r+1)+ \tau (q+1)\tau (p+1, r+1)}{\tau
(r+1)\tau (p+1, q+1)}
$$
is a periodic function of $r$ with period $1$ and an arbitrary function of
$p,q$. Because no special periodicity is implied, we set this function to
be $r$-independent.
Therefore, we arrive at the relation
$$
\tau (p+1)\tau (q+1, r+1)+ \tau (q+1)\tau (p+1, r+1)= h(2p, 2q)\tau
(r+1)\tau (p+1, q+1)\,,
$$
where $h$ can be an arbitrary function of $p$ and $q$. In the
original variables this equation reads
$$
T(a+1)T(a-1)+T(s+1)T(s-1)=h( u\! -\! s\! -\! a, u\! +\! s\! +\! a)\,
T(u+1)T(u-1)\,.
$$From the boundary conditions (\ref{T3}) at $a=0$ or $s=0$
it follows that $h=1$ and we obtain the Hirota equation
(\ref{HIROTA}). Note that the parameter $z$ entering the linear
problems disappears from the non-linear equation. In fact this is
clear from the very beginning because $z$ can be eliminated from
equations (\ref{LP3}) by the transformation $\psi \to z^{p+q+r}\psi$.
Nevertheless, we keep this parameter because it will be important
in what follows.

An advantage of the ``chiral" variables is their separation in
the linear problems: the first problem does not involve $q$ while
the second one does not involve $p$. However, in contrast to the
``laboratory" variables $a,s,u$, they have no immediate physical
meaning. Coming back to the ``laboratory" variables, we set $\psi
(p,q,r)=\Psi (q+r, \, -p-r, \, p+q)$ and rewrite the linear problems
(\ref{LP3}) in the form
\begin{equation}\label{LP4}
\begin{array}{l}
\displaystyle{\Psi (a,s,u) + z\frac{T(a\! -\! 1, s\! +\! 1, u) T(a, s\! -\! 1,
u\! +\! 1)}{T(a, s, u)\, T (a\! -\! 1, s, u\! +\! 1)}\, \Psi
(a\! -\! 1, s\! +\! 1, u)=
\Psi (a\!\! -\!\! 1,s, u\!\! +\!\! 1)}\\ \\
\displaystyle{\Psi (a,s,u)  - z\frac{T(a\! -\! 1, s\! +\!
1, u)T(a\! +\! 1, s, u\! +\! 1)}{T(a, s, u)\, T (a,s\! +\! 1, u\!
+\! 1) }\, \Psi (a\! -\! 1, s\! +\! 1, u)=\Psi (a, s\!\! +\!\! 1,
u\!\! +\!\! 1)}\,.
\end{array}
\end{equation}
In general, compatibility of linear problems implies existence of a
continuous family of common solutions. As we have seen, the
structure of the coefficient functions in our case is such that the
compatibility is equivalent to the existence of {\it at least one}
common solution (see \cite{Krichever06}, where this fact was pointed
out in another context).

Because the $T$-functions can vanish identically at some $a,s$, we
eliminate the denominators by passing to the new auxiliary function
$F=T\Psi$, in terms of which we have
\begin{equation}\label{LP5}
\begin{array}{l}
T(a\!\! -\!\! 1, s, u \!\! +\!\! 1)F(a, s, u)  + zT(a, s\!
-\! 1, u\!\! +\!\! 1)F(a\!\! -\!\! 1, s\!\! +\!\! 1, u)
=T(a, s, u)F(a\!\! -\!\! 1,s,u \!\! +\!\! 1)\\ \\
T(a,s\!\! +\!\! 1,u \!\! +\!\! 1)F(a, s, u)  - zT(a\!\! +\!\!
1, s, u\!\! +\!\! 1)F(a\!\! - \!\! 1, s\!\! +\!\! 1, u) =
T(a, s, u)F(a,s\!\! +\!\!
1,u \!\! +\!\! 1)\,.
\end{array}
\end{equation}
Note that the second equation can be obtained from the first one by
the transformation $T(a,s,u)\longrightarrow (-1)^{as}T(-s, -a, u)$
(and the same for $F$) which leaves the Hirota equation invariant.
However, the Hirota equation
written for the function $\tilde T(a,s,u)=
(-1)^{\frac{1}{2}(a^2 +s^2)}T(a,s,u)$
is form-invariant with respect to a larger symmetry group consisting
of any permutations and changing signs of the variables $a,s,u$
but the system of the linear problems (\ref{LP5}) is not. In fact
the symmetry is realized in an implicit way. To make it explicit,
we write the pair of equations (\ref{LP5}) in a matrix form,
\begin{equation}\label{LP51}
\begin{array}{c}
\left ( \begin{array}{cc}
T(a\! -\! 1, s, u)& zT(a, s\! -\! 1, u)\\ &\\
T(a, s\! +\! 1, u) & -zT(a\! +\! 1, s, u)
\end{array} \right )
\left ( \begin{array}{c} F(a,\, s,\, u\! -\! 1)\\ \\ F(a\! -\! 1,s\!
+\! 1,u\! -\! 1)
\end{array} \right )\\ \\
= \,\,\, T(a,s,u\! -\! 1)
\left ( \begin{array}{c}
F(a \! -\! 1, s, u)\\ \\
F(a, s\! +\! 1, u)
\end{array} \right ),\end{array}
\end{equation}
and
multiply both sides by the matrix inverse to the
one in the left hand side. Using the $TT$-relation,
we get another pair of linear problems,
\begin{equation}\label{LP52}
\begin{array}{l}
T(a\!\! +\!\! 1, s\!\! +\!\! 1, u)F(a, s, u)  - zT(a\!\! +\!\!
1, s, u\!\! +\!\! 1)F(a, s\!\! +\!\! 1, u\!\! -\!\! 1)
=T(a, s, u)F(a\!\! +\!\! 1,s\!\! +\!\! 1, u)\\ \\
T(a, s, u\!\! +\!\! 1) F(a, s, u\!\! -\!\! 1) - T(a,\,
s\!\! -\!\! 1, u)\, F(a, s\!\! +\!\! 1, u)  = T(a\!\! +\!\! 1,\, s, u)
 F(a\!\! -\!\! 1, s, u)
\end{array}
\end{equation}
which are equivalent to (and thus compatible with) the pair
(\ref{LP5}) by construction. The set of four linear problems
(\ref{LP5}), (\ref{LP52}) possesses the required symmetry.
The Hirota equation can be derived as a compatibility condition for any
two linear problems of these four, and the other two hold
automatically. The four linear equations
can be combined into a single matrix equation
\beq\label{LP531}
\TT (a,s,u)
\left (
\begin{array}{c}
F(a\! -\! 1,\, s,\, u) \\ \\ F(a,\, s\! +\! 1,\, u) \\ \\
F(a,\, s,\, u\! -\! 1) \\ \\ z F(a\!\! -\!\! 1, s\!\! +\!\! 1,
u\!\! -\!\! 1)
\end{array}
\right )=0 ,
\eeq
where $\TT (a,s,u)$ is the antisymmetric matrix
\begin{equation}\label{LP53}
\TT (a,s,u)=\left (
\begin{array}{cccc}
0 & T(a,s,u\! -\! 1) & - T(a,s\! +\! 1,u) & T(a\! +\! 1,s,u)
\\ &&&\\
-T(a,s,u\! -\! 1) & 0 & T(a\! -\! 1, s,u) & T(a,s\! -\! 1,u)
\\ &&& \\
T(a,s\! +\! 1,u) & -T(a\! -\! 1,s,u) & 0 & -T(a,s,u\! +\! 1)
\\ &&& \\
-T(a\! +\! 1,s,u) & -T(a,s\! -\! 1, u) & T(a,s,u\! +\! 1) & 0
\end{array} \right ).
\end{equation}
The Hirota equation implies that its
determinant vanishes and rank of this matrix equals 2.
The symmetric form of the linear
problems for the Hirota equation was suggested in
\cite{Shinzawa-1}.
 For more
details on the linear problems and their
symmetries see \cite{SS,Z2,Shinzawa-1,Shinzawa-2}.

\section{B\"acklund transformations}

There is a remarkable duality  between $T(a,s,u)$ and $F(a, s, u)$
\cite{SS,KLWZ}: one can exchange their roles
and treat eqs. (\ref{LP5}) as an over-determined system of linear
problems for the function $T$ with coefficients $F$. Their
compatibility condition is the same Hirota equation for $F$:
\beq\label{LP6}
F(a,s,u+1)F(a, s,u-1)= F(a, s+1,u)F(a, s-1,u)+
F(a+1,s,u)F(a-1,s,u)\,.
\eeq
 We thus conclude that any solution to the linear problems
(\ref{LP5}), where the $T$-function obeys the Hirota equation,
provides an (auto) B\"acklund transformation, i.e., a transformation
that sends a solution of the nonlinear integrable equation to
another solution of the same equation.

\begin{figure}[t]
   \centering
        \includegraphics[angle=-00,scale=0.6]{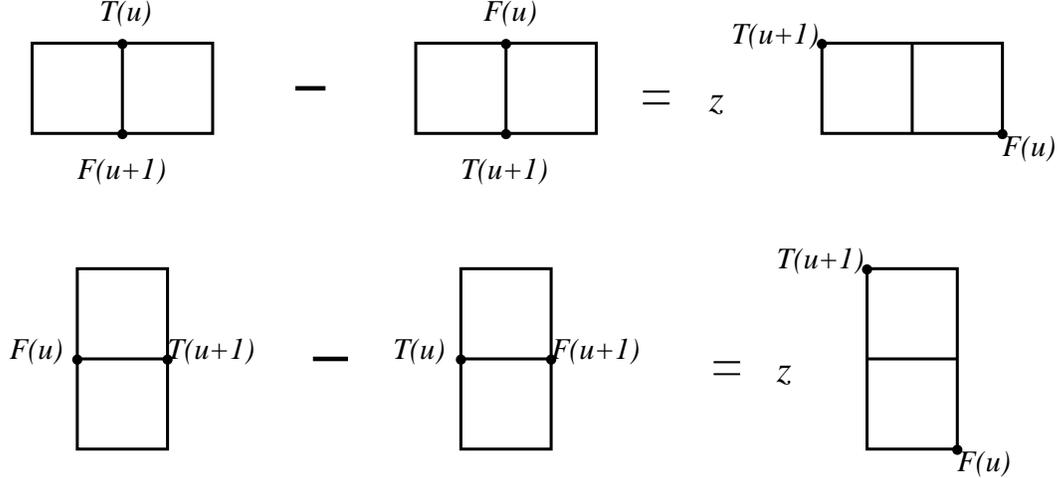}
        \caption{\it  The graphical representation of equations
(\ref{LINPR1}) in the $(a,s)$-lattice. Here $a$ and $s$ coordinates
correspond to the vertical and horizontal axis respectively.}
    \label{fig:BT1}
\end{figure}

Let us rewrite the linear problems (\ref{LP5}) changing the order of
the terms and shifting the variables:
\begin{equation}
\label{LINPR1}
\begin{array}{l}
  T(a\!\! +\!\! 1,s,u)F(a,s,u\!\! +\!\! 1)-
  T(a,s,u\!\! +\!\! 1)F(a\!\! +\!\! 1,s,u) =
  zT(a\!\! +\!\! 1,s\!\! -\!\! 1,u\!\! +\!\! 1)
  F(a,s\!\! +\!\! 1,u)\\ \\
   T(a,s\!\! +\!\! 1,u\!\! +\!\! 1)F(a,s,u)-
   T(a,s,u)F(a,s\!\! +\!\! 1,u\!\! +\!\! 1) =
   zT(a\!\! +\!\! 1,s,u\!\! +\!\! 1)
   F(a\!\! -\!\! 1,s\!\! +\!\! 1,u).
   \end{array}
\end{equation}
These equations are graphically represented in Fig.~\ref{fig:BT1} in
the $(a,s)$-plane. They constitute the B\"acklund transformation
$T\to F$ in the bilinear form \cite{Hirota,SS}. Given a family
of polynomials $T(a,s,u)$ obeying the Hirota equation, one may
pose the problem
of finding polynomial solutions to equations (\ref{LINPR1}).
It is easy to see  that
these equations are not compatible with the
boundary conditions for $F(a,s,u)$ and $T(a,s,u)$ of the ``fat hook"
type with the same $K$ and $M$. Indeed, applying these equations
in the corner
point of the interior boundary, one sees that if
$K,M$ for $T$ and $F$ are the same, then
the boundary values
must vanish identically. However, it is straightforward to verify
that equations (\ref{LINPR1}) are compatible with the boundary
conditions of the following two types:
\begin{equation}
\label{BCONDF1}
\begin{array}{l}
  F(a,s,u)=0\quad  {\rm if:}\\ \\
    \mbox{(i)} \;\; a<0
  \quad {\rm or} \quad \mbox{(ii)} \;\; a>0 \;\; {\rm and} \;\;
  s<0, \quad {\rm or} \quad \mbox{(iii)} \;\; a>K\! -\! 1
  \;\; {\rm and}\;\;  s>M,
  \end{array}
\end{equation}
or
\begin{equation}
\label{BCONDF1a}
\begin{array}{l}
  F(a,s,u)=0\quad  {\rm if:}\\ \\
    \mbox{(i)} \;\; a<0
  \quad {\rm or} \quad \mbox{(ii)} \;\; a>0 \;\; {\rm and} \;\;
  s<0, \quad {\rm or} \quad \mbox{(iii)} \;\; a>K
  \;\; {\rm and}\;\;  s>M+1\,.
  \end{array}
\end{equation}
They are again of the ``fat hook" type but with the shifts $K\to
K-1$ or $M\to M+1$. We refer to the corresponding transformations as
$\mbox{BT}_{1}^{-}$ and $\mbox{BT}_{2}^{+}$:
$F(a,s,u)=\mbox{BT}_{1}^{-}(T(a,s,u))$ for (\ref{BCONDF1}) and
$F(a,s,u)=\mbox{BT}_{2}^{+}(T(a,s,u))$ for (\ref{BCONDF1a}).
They depend on the parameter $z$. The same formulas (\ref{LINPR1})
define inverse transformations
$\mbox{BT}_{1}^{+}=(\mbox{BT}_{1}^{-})^{-1}$ and
$\mbox{BT}_{2}^{-}=(\mbox{BT}_{2}^{+})^{-1}$ if one treats them
as linear equations for $T$ with given $F$.
In a more explicit way, the transformations $\mbox{BT}_{1}^{\pm}$,
$\mbox{BT}_{2}^{\pm}$ are defined by formulas
(\ref{LP531b}), (\ref{LP531a})
below.

Repeating these transformations several times,
we arrive at the hierarchy of functions $T_{k,m}(a,s,u)$
($k=0,1, \ldots , K$, $m=0,1, \ldots , M$) such that:
\begin{itemize}
\item[a)]
They obey the Hirota equation in $a,s,u$ for any $k,m$;
\item[b)]
They are connected by the B\"acklund transformations
\beq\label{Backlund}
\begin{array}{l}
T_{k-1,m}(a,s,u)=\mbox{BT}_{1}^{-}(T_{k,m}(a,s,u)),\\ \\
T_{k,m-1}(a,s,u)=\mbox{BT}_{2}^{-}(T_{k,m}(a,s,u));
\end{array}
\eeq
\item[c)]
At $k=K$, $m=M$ we have
$$
T_{K,M}(a,s,u)=T(a,s,u)\,.
$$
\end{itemize}
For the irreducible polynomial solutions this list
should be supplemented by the ``initial condition"
$T_{0,0}(a,s,u)=1$ at $a=0$ or at $s=0$, $a>0$ and $T_{0,0}(a,s,u)=0$
otherwise. The levels of the hierarchy are labeled by the
pair of numbers $k,m$. The lowest level is $0,0$ while
the highest one is $K,M$.
We shall say that the $T$-functions $T_{k,m}(a,s,u)$
belong to the level $k,m$.

It is important to note that the parameter $z$ involved in the
definition of the B\"acklund transformations can be different
for transformations of the two types
introduced above as well as for successive
transformations of the same type. We choose these parameters
to be eigenvalues of the matrix $g$: $z=x_k$ for
the transition $(k,m)\to (k-1,m)$ and $z=y_m$ for
$(k,m)\to (k,m-1)$.
More precisely, the transformation $\mbox{BT}_{1}^{-}$ is written
in the form
\begin{equation}
\label{LINPRT1}
\begin{array}{c}
  T_{k,m}(a+1,s,u)T_{k-1,m}(a,s,u+1)-T_{k,m}(a,s,u+1)T_{k-1,m}(a+1,s,u)
\\ \\
=\,\, x_k \, T_{k,m}(a+1,s-1,u+1)T_{k-1,m}(a,s+1,u), \\ \\
  T_{k,m}(a,s+1,u+1)T_{k-1,m}(a,s,u)-T_{k,m}(a,s,u)T_{k-1,m}(a,s+1,u+1)
\\ \\
=\,\,  x_k \, T_{k,m}(a+1,s,u+1)T_{k-1,m}(a-1,s+1,u),
\end{array}
\end{equation}
while $\mbox{BT}_{2}^{-}$ in the form
\begin{equation}
\label{LINPRT2}
\begin{array}{c}
  T_{k,m-1}(a+1,s,u)T_{k,m}(a,s,u+1)-T_{k,m-1}(a,s,u+1)T_{k,m}(a+1,s,u)
\\ \\
=\,\, y_m \, T_{k,m-1}(a+1,s-1,u+1)T_{k,m}(a,s+1,u), \\ \\
   T_{k,m-1}(a,s+1,u+1)T_{k,m}(a,s,u)-T_{k,m-1}(a,s,u)T_{k,m}(a,s+1,u+1)
\\ \\
=\,\, y_m \,   T_{k,m-1}(a+1,s,u+1)T_{k,m}(a-1,s+1,u)
\end{array}
\end{equation}
($k=1,\dots,K$, $m=1,\dots , M$). If one ignores the $u$-dependence
(which disappears in the $u\to \infty$ limit), then these formulas
become the bilinear relations between characters mentioned
in Section 3 (see equations (\ref{ch8}) and comments after them).
It is easy to notice that each of the equations in
(\ref{LINPRT1}), (\ref{LINPRT2}) is actually
a dynamical equation for a function
of three variables rather than five.
For example, the first equation in (\ref{LINPRT1}) acts in the
subspaces $m=\mbox{const}$ and $u+s+a=\mbox{const}$.
Upon restriction to the corresponding three-dimensional hyperplanes
in the linear space with coordinates $a,s,u,k,m$, each of these
equations
can be put in the standard Hirota form by a linear
change of variables.

For completeness, let us give a symmetric description of the
B\"acklund transformations through the matrix equations
of the form (\ref{LP531}). For the direct and inverse
transformations we need to introduce two antisymmetric
$4\times 4$ matrices $\TT ^{(\pm 1)} (a,s,u)$
of the type (\ref{LP53}):
\beq\label{LP54}
\TT ^{(\varepsilon )} (a,s,u)=\left (
\begin{array}{cccc}
0 & T(a,s,u\! -\! \varepsilon ) & -
T(a,s\! +\! \varepsilon ,u) & T(a\! +\! \varepsilon ,s,u)
\\ &&&\\
-T(a,s,u\! -\! \varepsilon ) & 0 &
T(a\! -\! \varepsilon , s,u) & T(a,s\! -\! \varepsilon ,u)
\\ &&& \\
T(a,s\! +\! \varepsilon ,u) & -
T(a\! -\! \varepsilon ,s,u) & 0 & -T(a,s,u\! +\! \varepsilon )
\\ &&& \\
-T(a\! +\! \varepsilon ,s,u) & -
T(a,s\! -\! \varepsilon , u) & T(a,s,u\! +\! \varepsilon ) & 0
\end{array} \right )
\eeq
where $\varepsilon =\pm 1$.
Then the transformations $T_{k-1,m}=\mbox{BT}_{1}^{-}(T_{k,m})$,
$T_{k,m+1}=\mbox{BT}_{2}^{+}(T_{k,m})$
are defined by the matrix equation
\beq\label{LP531b}
\TT^{(+1)}_{k,m}(a,s,u)
\left (
\begin{array}{cc}
T_{k-1,m}(a\! -\! 1,\, s,\, u)& T_{k,m+1}(a\! -\! 1,\, s,\, u)
\\ & \\ T_{k-1,m}(a,\, s\! +\! 1,\, u)&
T_{k,m+1}(a,\, s\! +\! 1,\, u)\\ & \\
T_{k-1,m}(a,\, s,\, u\! -\! 1)& T_{k,m+1}(a,\, s,\, u\! -\! 1)\\ &\\
x_k T_{k-1,m}(a\!\! -\!\! 1, s\!\! +\!\! 1, u\!\! -\!\! 1)&
y_{m+1} T_{k,m+1}(a\!\! -\!\! 1, s\!\! +\!\! 1, u\!\! -\!\! 1)
\end{array}
\right )=0 ,
\eeq
and the inverse transformations $T_{k+1,m}=\mbox{BT}_{1}^{+}(T_{k,m})$,
$T_{k,m-1}=\mbox{BT}_{2}^{-}(T_{k,m})$ are defined by the
equation
\beq\label{LP531a}
\TT^{(-1)}_{k,m}(a,s,u)
\left (
\begin{array}{cc}
T_{k+1,m}(a\! +\! 1,\, s,\, u)& T_{k,m-1}(a\! +\! 1,\, s,\, u)
\\ & \\ T_{k+1,m}(a,\, s\! -\! 1,\, u)&
T_{k,m-1}(a,\, s\! -\! 1,\, u)\\ & \\
T_{k+1,m}(a,\, s,\, u\! +\! 1)& T_{k,m-1}(a,\, s,\, u\! +\! 1)\\ &\\
x_{k+1} T_{k+1,m}(a\!\! +\!\! 1, s\!\! -\!\! 1, u\!\! +\!\! 1)&
y_{m} T_{k,m-1}(a\!\! +\!\! 1, s\!\! -\!\! 1, u\!\! +\!\! 1)
\end{array}
\right )=0 ,
\eeq
where $\TT^{(\pm )}_{k,m} (a,s,u)$ are the matrices (\ref{LP54}) with entries
at the level $k,m$.

Moreover, a careful analysis of the equations
(\ref{LINPRT1})-(\ref{LP531a}) shows that boundary values of the
$T$-functions at each level $k,m$ are subject to the same relations
as at the highest level. On the exterior boundaries, the
$T$-functions have the specific form similar to (\ref{T3}), i.e.,
$T_{k,m}(0,s,u)$ is a function of $u-s$ while $T_{k,m}(a,0,u)$ is
the same function of $u+a$. Let us introduce the special notation
for them:
\beq\label{Q}
T_{k,m}(0,s,u)=Q_{k,m}(u-s)\,, \quad T_{k,m}(a,0,u)=Q_{k,m}(u+a)\,.
\eeq
The polynomials $Q_{k,m}(u)$ play a very important role.
They will be identified with eigenvalues of Baxter's $Q$-operators.
The polynomial $Q_{K,M}(u)=\phi (u)$ is a fixed input data which
determines the model. The polynomials $Q_{k,m}(u)$ at lower levels
are to be found in the course of solution.
In analogy with (\ref{F13}), we fix their
highest coefficients to be $1$.
Applying (\ref{LINPRT1}), (\ref{LINPRT2}) on the interior
boundaries, we conclude that
if (\ref{IDENT}) is valid at the highest level, then
$T$-functions on the interior boundaries at each level
$k,m$ are connected by
a similar relation:
\begin{equation} \label{IDENT1}
  T_{k,m}(k,m+n,u)=(-1)^{nm} (\mbox{sdet}\, g_{k,m})^n \,
  T_{k,m}(k+n,m,u),\quad  n\ge 0.
\end{equation}
The matrix $g_{k,m}\in GL(k|m)$ is defined in (\ref{gkmdiag})
and $\mbox{sdet}\, g_{k,m}=x_1 \ldots x_k/(y_1 \ldots y_m)$.

We see now that by applying $\mbox{BT}_{1}^{-}$ and
$\mbox{BT}_{2}^{-}$ to a solution of the
Hirota equation in the domain ${\sf H}(K|M)$
with the boundary conditions (\ref{BCOND1}), we can
successively transform it to the trivial solution in the degenerate domain
${\sf H}(0|0)$. This ``undressing
procedure" allows one to construct solutions to the original
problem, as it will be shown below.

\section{Recurrence relations for the operator generating series}

The B\"acklund transformations from the previous section
can be reformulated in the operator form as recurrence relations
for difference operators of infinite order.
Let us consider the following operator:
\beq\label{R1}
{\cal W}(u)=\sum_{s\geq 0}\frac{T(1,s, u+s-1)}{\phi
(u)}\, e^{2s\p_u}
\eeq
(the common denominator is introduced for normalization).
It serves as a non-commutative generating
series for the $T$-functions $T(1,s,u)$. Similar objects can be
introduced at any level $k,m$:
\beq\label{R2}
{\cal W}_{k,m}(u)=\sum_{s\geq 0}\frac{T_{k,m}(1,s, u+s-1)}{Q_{k,m} (u)}\,
e^{2s\p_u}\,.
\eeq
It is the generating series for the functions $T_{k,m}(1,s,u)$.
Clearly, ${\cal W}_{0,0}(u)=1$ (recall that $T_{0,0}(1,s,u)=0$ unless
$s=0$ and $T_{0,0}(1,0,u)=Q_{0,0}(u+1)=1$).
We also note that the series formally inverse to (\ref{R2})
generates the $T$-functions $T_{k,m}(a,1,u)$:
\beq\label{R2a}
{\cal W}^{-1}_{k,m}(u)=
\sum_{a\geq 0}(-1)^a \,
e^{2a\p_u}\, \frac{T_{k,m}(a,1, u-a-1)}{Q_{k,m} (u-2)} \,.
\eeq
For the proof, see \cite{KSZ}.

Set
\beq\label{R3}
X_{k,m}(u)=x_k\, \frac{Q_{k,m}(u\!
+\! 2)\, Q_{k\! -\! 1,m}(u\! -\! 2)}{Q_{k,m} (u)\,\, Q_{k-1, m}(u)}\,,
\eeq
\beq\label{R4}
Y_{k,m}(u)=y_m\, \frac{Q_{k,m\! -\! 1}(u\! +\! 2)\,
Q_{k,m}(u \! -\! 2)}{Q_{k,m-1} (u)\,\, Q_{k, m}(u)}\,.
\eeq
Using the linear problems at $a=0$, it is a straightforward calculation
to prove the following recurrence relations for the operators
${\cal W}_{k,m}(u)$:
\beq\label{R5}
\begin{array}{l}
{\cal W}_{k-1,\, m}(u)=\left (1-X_{k,m}(u)e^{2\p_u}\right )
{\cal W}_{k,m}(u)\,, \\ \\
{\cal W}_{k,\, m+1}(u)=\left (1-Y_{k,m+1}(u)e^{2\p_u}\right )
{\cal W}_{k,m}(u)\,.
\end{array}
\eeq
These formulas are operator (and $u$-dependent)
analogs of (\ref{rec}).
The shift operator ${\bf t}=e^{2\p_u}$ plays the role of the
variable $t$ while the variable $u$ is absent in (\ref{rec}).
Notice also that $X_{k,m}(u)$, $Y_{k,m}(u)$ turn into $x_k$,
$y_m$ in the limit $u\to \infty$.

Here we present some details of the proof
(see also \cite{KSZ}, where a slightly different
version of the recurrence relations is proved).
Consider the first relation.
We have:
\beq\label{A1}
{\cal W}_{k-1,m}(u)-{\cal W}_{k,m}(u)=
\sum_{s\geq 0}\left [
\frac{T_{k-1,m}(1,s,u+s-1)}{Q_{k-1,m}(u)}-
\frac{T_{k,m}(1,s,u+s-1)}{Q_{k,m}(u)}\right ]
e^{2s\p_u} \,.
\eeq
To transform the expression in the square brackets, we rewrite
the first equation in (\ref{LINPRT1}) at $a=0$ in the form
$$
\begin{array}{c}
\displaystyle{
\frac{T_{k,m}(1,s,u+s-1)}{Q_{k,m}(u)} \, -\,
\frac{T_{k-1,m}(1,s,u+s-1)}{Q_{k-1,m}(u)}}
\\ \\ =\,\,\,\,
\displaystyle{
x_k \, \frac{Q_{k,m}(u+2)Q_{k-1,m}(u-2)}{Q_{k,m}(u)Q_{k-1,m}(u)}
\, \frac{T_{k,m}(1, s\! -\! 1, u\! +\! s)}{Q_{k,m}(u+2)}}
\end{array}
$$
and continue the equality:
\beq\label{A2}
{\cal W}_{k-1,m}(u)-{\cal W}_{k,m}(u)=-\, X_{k,m}(u)\sum_{s\geq 0}\,
\frac{T_{k,m}(1, s\! -\! 1, u\! +\! s)}{Q_{k,m}(u+2)}\,\, e^{2s\p_u}\,.
\eeq
Because $T_{k,m}(1,-1,u)=0$, the sum in right hand side can be written
in the form
$$
\sum_{s\geq 0}\, \frac{T_{k,m}(1, s\! -
\! 1, u\! +\! s)}{Q_{k,m}(u+2)}\,\, e^{2s\p_u}=
e^{2\p_u}\sum_{s\geq 0}\, \frac{T_{k,m}(1, s,u+s-1)}{Q_{k,m}(u)}
=e^{2\p_u}{\cal W}_{k,m}(u)\,,
$$
and the first recurrence relation is proved.
The proof of the second one is completely similar.

\section{Factorization formulas and $TQ$-relations}

The recurrence relations established in the previous
section allow one to represent the operator generating
series (\ref{R1}) in a closed factorized form, where each factor
contains the $Q$-functions only. Namely, ${\cal W}_{K,M}(u)$
can be obtained as a result of successive application of
the recurrence relations (\ref{R5}) to ${\cal W}_{0,0}(u)=1$.
In this way, moving first in the $m$-direction from
$(0,0)$ to $(0,M)$ and then in the $k$-direction
from $(0,M)$ to $(K,M)$, we get:
\beq\label{B1}
{\cal W}_{K,M}(u)=\prod_{K\geq k\geq 1}^{\leftarrow}
\left (1-X_{k,M}(u)e^{2\p_u}\right )^{-1} \cdot
\prod_{M\geq m \geq 1}^{\leftarrow}
\left (1-Y_{0,m}(u)e^{2\p_u}\right )
\eeq
where the ordered product is defined as
$$
\prod_{J\geq i\geq I}^{\leftarrow}A_i =
A_J \, A_{J-1} \, \ldots \, A_{I+1}\, A_I\,.
$$
Applying the recurrence relations in the different order
($k$-direction first, $m$-direction next), we arrive at
a different but equivalent representation:
\beq\label{B2}
{\cal W}_{K,M}(u)=\prod_{M\geq m\geq 1}^{\leftarrow}
\left (1-Y_{K,m}(u)e^{2\p_u}\right ) \cdot
\prod_{K\geq k \geq 1}^{\leftarrow}
\left (1-X_{k,0}(u)e^{2\p_u}\right )^{-1}.
\eeq

In fact one can apply the relations (\ref{R5}) in any other order
determined by a chosen zigzag path from the point $(0,0)$ to the point
$(K,M)$. Each step in the $k$-direction,
$(k,m)\to (k+1,m)$, brings the factor
$\left (1-X_{k+1, m}(u)e^{2\p_u}\right )^{-1}$ while each step
in the $m$-direction, $(k,m)\to (k, m+1)$, brings the factor $\left
(1-Y_{k,m+1}(u)e^{2\p_u}\right )$ which are to be multiplied
according to the order of the steps. This yields many other ways to
factorize the operator ${\cal W}_{K,M}(u)$. Their equivalence
follows from the compatibility of the recurrence relations
(\ref{R5}) which is expressed by the discrete ``zero curvature"
condition
\beq\label{zc}
\left (1-Y_{k-1,m+1}(u)e^{2\p_u}\right )
 \left (1-X_{k,m}(u)e^{2\p_u}\right )=
\left (1-X_{k,m+1}(u)e^{2\p_u}\right )
\left (1-Y_{k,m+1}(u)e^{2\p_u}\right )
\eeq
on the $(k,m)$-lattice. The two sides of this equality
correspond to two different ways to obtain ${\cal W}_{k-1, m+1}(u)$
from ${\cal W}_{k, m}(u)$.

The equalities (\ref{B1}) and (\ref{B2}) as well as the similar
equalities with different orderings are generalized Baxter's
$TQ$-relations in a generating form. Equating coefficients in front
of different powers of the operator $e^{2\p_u}$, one obtains
expressions for the $T$-functions $T(1,s,u)=T_{K,M}(1,s,u)$ through
$X_{k,m}$, $Y_{k,m}$ and thus through the $Q$-functions $Q_{k,m}(u)$
with $1\leq k \leq K$, $1\leq m \leq M$. For example, the simplest
$TQ$-relation contained in (\ref{B1}) has the form
\beq\label{B3}
\frac{T_{K,M}(1,1,u)}{Q_{K,M}(u)}=\sum_{k=1}^{K}X_{k,M}(u)-
\sum_{m=1}^{M}Y_{0,m}(u)\,,
\eeq
where $X_{k,M}(u)$, $Y_{0,m}(u)$
are to be expressed through the $Q$-functions according to
(\ref{R3}), (\ref{R4}).
The zeros of the latter are to be constrained by the system of
Bethe equations derived in the next section.

\section{$QQ$-relation and Bethe equations}

Our starting point in this section is the discrete
zero curvature condition (\ref{zc}). Comparing coefficients
in front of different powers of the shift operator, we note
that
$$
Y_{k-1,m+1}(u)X_{k,m}(u+2)=X_{k,m+1}(u)Y_{k,m+1}(u+2)
$$
holds identically, and thus get the only non-trivial relation
$$
Y_{k-1,m+1}(u)+X_{k,m}(u)=X_{k,m+1}(u)+Y_{k,m+1}(u)\,,
$$
which after the substitution
(\ref{R3}), (\ref{R4}) becomes a functional equation for
the $Q$-functions.
As a simple calculation shows, it is equivalent
to the following bilinear equation
(the ``$QQ$-relation" \cite{KSZ}):
\begin{equation} \label{QHIROTA}
\begin{array}{c}
  x_{k}Q_{k-1,m-1}(u)Q_{k,m}(u+2)- y_{m}Q_{k,m}(u)Q_{k-1,m-1}(u+2)
  \\ \\ =\,\,\,\,
  (x_{k}-y_{m})\, Q_{k-1,m}(u)Q_{k,m-1}(u+2)
  \end{array}
\end{equation}
for the polynomial functions
\beq\label{R6}
Q_{k,m}(u)=\prod_{j=1}^{N_{k,m}} (u-u_{j}^{(k,m)})\,.
\eeq

Let us remark that the $QQ$-relation (\ref{QHIROTA}) can be
recast to the standard form of the Hirota bilinear difference
equation in ``chiral" variables $u,m, -k$ by passing to the function
\beq\label{calQ}
{\cal Q}_{k,m}(u)=a_{k,m}e^{(\beta_k +\gamma_m )u}\, Q_{k,m}(u)\,,
\eeq
where the new parameters $\beta_k$ and $\gamma_m$ are related to
the $x_k$, $y_m$ by the formulas
\beq\label{R8}
x_k = e^{2(\beta_k -\beta_{k-1})}\,,
\quad
y_m = e^{2(\gamma_{m-1}-\gamma_m )}\,.
\eeq
They are fixed uniquely by putting $\beta_0 = \gamma_0 =0$.
This transformation eliminates the coefficients $x_k$, $y_m$ in
(\ref{QHIROTA}) (as well as in (\ref{R3}), (\ref{R4})) and, with
a proper choice of $a_{k,m}$, the $QQ$-relation acquires the
coefficient-free form
\begin{equation} \label{QHIROTA1}
  {\cal Q}_{k-1,m-1}(u){\cal Q}_{k,m}(u+2)-
  {\cal Q}_{k,m}(u){\cal Q}_{k-1,m-1}(u+2)=
  {\cal Q}_{k-1,m}(u){\cal Q}_{k,m-1}(u+2)
\end{equation}
suggested in \cite{KSZ}.

The $QQ$-relation (\ref{QHIROTA}) provides the easiest
and the most transparent way to derive Bethe equations
for roots of the polynomials $Q_{k,m}(u)$. Putting
$u$ in (\ref{QHIROTA}) successively equal to the roots
of each $Q$-function entering the equation, one obtains
a number of relations which, after some rearranging, can be
written in the form

\begin{equation}\label{Bethe1}
\frac{Q_{k-1,m}\left
(u_{j}^{(k,m)}\right ) Q_{k,m}\left (u_{j}^{(k,m)}-2\right )
Q_{k+1,m}\left (u_{j}^{(k,m)}+2\right )}{ Q_{k-1,m}\left
(u_{j}^{(k,m)}-2\right ) Q_{k,m}\left (u_{j}^{(k,m)}+2\right )
Q_{k+1,m}\left (u_{j}^{(k,m)}\right )}=\, -\, \frac{x_k}{x_{k+1}}\,,
\end{equation}

\vspace{3mm}

\begin{equation}\label{Bethe2}
\frac{Q_{k,m+1}\left
(u_{j}^{(k,m)}\right ) Q_{k,m}\left (u_{j}^{(k,m)}-2\right )
Q_{k,m-1}\left (u_{j}^{(k,m)}+2\right )}{ Q_{k,m+1}\left
(u_{j}^{(k,m)}-2\right ) Q_{k,m}\left (u_{j}^{(k,m)}+2\right )
Q_{k,m-1}\left (u_{j}^{(k,m)}\right )}=\, -\, \frac{y_{m+1}}{y_m}\,,
\end{equation}

\vspace{3mm}

\begin{equation}\label{Bethe5}
\frac{Q_{k+1,m}\left
(u_{j}^{(k,m)}\right ) Q_{k,m-1}\left (u_{j}^{(k,m)}+2\right )}{
Q_{k+1,m}\left (u_{j}^{(k,m)}+2\right ) Q_{k,m-1}\left
(u_{j}^{(k,m)}\right )}=\,  \frac{x_{k+1}}{y_m}\,,
\end{equation}

\vspace{3mm}

\begin{equation}\label{Bethe6}
\frac{Q_{k,m+1}\left
(u_{j}^{(k,m)}\right ) Q_{k-1,m}\left (u_{j}^{(k,m)}-2\right )}{
Q_{k,m+1}\left (u_{j}^{(k,m)}-2\right ) Q_{k-1,m}\left
(u_{j}^{(k,m)}\right )}=\,  \frac{y_{m+1}}{x_k}\,.
\end{equation}

\noindent
They hold inside the
$K\times M$ rectangle
in the $(k,m)$-lattice and serve as elementary building blocks
for systems of Bethe equations. Each such system corresponds to
a zigzag ``undressing" path from $(K,M)$ to $(0,0)$. On the parts
of the path $(k+1,m)\to (k,m)\to (k-1, m)$,
$(k,m+1)\to (k,m)\to (k, m-1)$,
$(k+1,m)\to (k,m)\to (k, m-1)$ and
$(k,m+1)\to (k,m)\to (k-1, m)$ one uses (\ref{Bethe1}), (\ref{Bethe2}),
(\ref{Bethe5}) and (\ref{Bethe6}) respectively.
These systems
are different but equivalent. For a more detailed
discussion on this point, see \cite{KSZ}.
As an example, we give here
the chain of the Bethe equations for the simplest path
$(K,M)\longrightarrow (0,M) \longrightarrow (0,0)$.
Moving from $(K,M)$ to $(0,M)$, we have the equations
\begin{equation}\label{Bethe1c}
\frac{Q_{k-1,M}\left (u_{j}^{(k,M)}\right )
Q_{k,M}\left (u_{j}^{(k,M)}-2\right )  Q_{k+1,M}\left
(u_{j}^{(k,M)}+2\right )} { Q_{k-1,M}\left
(u_{j}^{(k,M)}-2\right ) Q_{k,M}\left (u_{j}^{(k,M)}+2\right
)  Q_{k+1,M}\left (u_{j}^{(k,M)}\right )}=\, -\,
\frac{x_k}{x_{k+1}}\,,
\end{equation}
where $k=1, \ldots , K-1$. They agree with the chain of Bethe
equations presented in \cite{KulResh2} for the bosonic case.
At the turning point, the equation is
\begin{equation}\label{Bethe5c}
\frac{Q_{1,M}\left (u_{j}^{(0,M)}\right )
Q_{0,M-1}\left (u_{j}^{(0,M)}+2\right )}{Q_{1,M}\left
(u_{j}^{(0,M)}+2\right ) Q_{0,M-1}\left (u_{j}^{(0,M)}\right
)}=\, \frac{x_1}{y_M}\,.
\end{equation}
Finally, moving from $(0,M)$ to $(0,0)$, we have the
equations
\begin{equation}\label{Bethe2c}
\frac{Q_{0,m+1}\left (u_{j}^{(0,m)}\right )
Q_{0,m}\left (u_{j}^{(0,m)}-2\right ) Q_{0,m-1}\left
(u_{j}^{(0,m)}+2\right )}{  Q_{0,m+1}\left
(u_{j}^{(0,m)}-2\right ) Q_{0,m}\left (u_{j}^{(0,m)}+2\right
) Q_{0,m-1}\left (u_{j}^{(0,m)}\right )}=\, -\,
\frac{y_{m+1}}{y_m}\,,
\end{equation}
where $m=1,2, \ldots , M-1$.

\section{Conclusion}

We have obtained a solution of the
$TT$-relation (the Hirota equation) obeying all the
required boundary and analytic conditions. The solution
is given by the determinant formula (\ref{BRdet}), where
the polynomials $T(1,s,u)$ are determined by the
expansion (\ref{R1}) of the factorized operator
(\ref{B1}). The coefficients of the latter are expressed
through the $Q$-functions via equations (\ref{R3}), (\ref{R4}),
where the roots of the polynomials $Q_{k,m}(u)$ are constrained by
Bethe equations which result from the bilinear
$QQ$-relation (\ref{QHIROTA}). It should be noted that the
solution is not unique. Given boundary conditions,
there is a finite set of solutions
corresponding to different quantum states of the generalized
spin chain.

We emphasize that this solution, typical for quantum
problems solvable by Bethe ansatz,
has been obtained by purely classical methods of the
theory of soliton equations on the lattice. The key role
is played by B\"acklund transformations for the Hirota
difference equation. From the viewpoint of
the theory of classical
soliton equations, our method consists in constructing a chain of
successive B\"acklund transformations which reduces the
problem to a trivial one. Each transformation of this chain
involves a continuous parameter (the classical
spectral parameter) which is identified with an eigenvalue
of the matrix which defines the twisted
boundary conditions in the quantum integrable model.

\section*{Acknowledgments}

The author is grateful to V.Kazakov and A.Sorin for numerous
discussions and collaboration in \cite{KSZ}.
He also thanks the organizers of the Workshop ``Classical and
quantum integrable systems" (Dubna, January 2007), where some
of these results were reported, for a kind
hospitality. This work was supported in part by
grant RFBR-06-01-92054-$\mbox{CE}_{a}$,
by grant INTAS 03-51-6346, by grant for support of
scientific schools NSh-8004.2006.2 and by the ANR project GIMP No.
ANR-05-BLAN-0029-01.


\begin{thebibliography}{99}

\bibitem{KLWZ}
I.~Krichever, O.~Lipan, P.~Wiegmann and A.~Zabrodin, {\it Quantum
integrable systems and elliptic solutions of classical discrete
nonlinear equations}, Commun. Math. Phys. {\bf 188} (1997) 267-304,
arXiv.org: hep-th/9604080.

\bibitem{Z1} A.~Zabrodin, {\it Discrete Hirota's equation in
quantum integrable models}, Int. J. Mod. Phys. B {\bf 11} (1997)
3125-3158, arXiv.org: hep-th/9610039;

\bibitem{Z3}
A.~Zabrodin, {\it Hirota
equation and Bethe ansatz}, Teor. Mat. Fys. {\bf 116} (1998) 54-100
(English translation: Theor. Math. Phys. {\bf 116} (1998) 782-819).

\bibitem{BR1} V.~Bazhanov and N.~Reshetikhin,
{\it Restricted solid-on-solid models connected with simply laced
algebras and conformal field theory}, J. Phys. A: Math. Gen. {\bf
23} (1990) 1477-1492.

\bibitem{KlumperPearce} A.~Kl\"umper and P.~Pearce,
{\it Conformal weights of RSOS lattice models and their fusion
hierarchies}, Physica {\bf A183} (1992) 304-350.



\bibitem{Kuniba-1} A.~Kuniba, T.~Nakanishi and J.~Suzuki,
{\it Functional relations in solvable lattice models I: Functional
relations and representation theory}, Int. J. Mod. Phys. {\bf A9}
(1994) 5215-5312, arXiv.org: hep-th/9309137.

\bibitem{Tsuboi-1} Z.~Tsuboi, {\it Analytic Bethe ansatz and
functional equations for Lie superalgebra $sl(r+1|s+1)$}, J. Phys.
A: Math. Gen. {\bf 30} (1997) 7975-7991.


\bibitem{Hirota} R.~Hirota, {\it Discrete analogue of a
generalized Toda equation}, Journ. of the Phys. Soc. of Japan, {\bf
50} (1981) 3785-3791.


\bibitem{KSZ} V.~Kazakov, A.~Sorin and A.~Zabrodin,
{\it Supersymmetric Bethe Ansatz and Baxter Equations from
Discrete Hirota Dynamics}, arXiv.org: hep-th/0703147.

\bibitem{KulSk}  P.~Kulish and E.~Sklyanin,
{\it On solutions of the Yang-Baxter equation}, Zap. Nauchn. Sem.
LOMI {\bf 95} (1980)  129-160; Engl. transl.: J. Soviet Math.,  {\bf
19} (1982) 1956.


\bibitem{Kulish}
P.~Kulish, {\it Integrable graded magnetics}, Zap. Nauchn. Sem. LOMI
{\bf 145} (1985),  140-163.


\bibitem{Kac} V.~Kac, {\it Lie superalgebras}, Adv. Math.
{\bf 26} (1977) 8-96; V.~Kac, Lecture Notes in Mathematics,
{\bf 676}, pp. 597-626, Springer-Verlag, New York, 1978.

\bibitem{Bars2}
A.~Baha Balantekin and I.~Bars, {\it  Dimension And Character
Formulas For Lie Supergroups}, J.\ Math.\ Phys.\  {\bf 22}, 1149
(1981).

\bibitem{BMR} I.~Bars, B.~Morel and H.~Ruegg, {\it Kac-Dynkin
diagrams and supertableaux}, J. Math. Phys. {\bf 24} (1983)
2253-2262.



\bibitem{FT} L.~Faddeev and L.~Takhtadjan, {\it Quantum
inverse scattering method and the $XYZ$ Heisenberg model}, Uspekhi
Mat. Nauk, {\bf 34:5} (1979) 13-63.



\bibitem{Faddeev}
L.~Faddeev {\it Algebraic Aspects of Bethe-Ansatz},
Int. J. Mod. Phys.
{\bf A10} (1995) 1845-1878, arXiv.org: hep-th/9404013.

\bibitem{book} N.~Bogoliubov, A.~Izergin and V.~Korepin,
{\it Quantum inverse scattering method and correlation functions},
Cambridge: Cambridge University Press, 1993.

\bibitem{DJM} E.~Date, M.~Jimbo and T.~Miwa, {\it Method for
generating discrete soliton equations I}, J. Phys. Soc. Japan {\bf
51} (1982) 4116-4127.

\bibitem{Z2} A.~Zabrodin, {\it Hirota's difference equations},
Teor. Mat. Fys. {\bf 113} (1997) 179-230 (English translation:
Theor. Math. Phys. {\bf 113} (1997) 1347-1392), arXiv.org:
solv-int/9704001.

\bibitem{Macdonald} I.~Macdonald, {\it Symmetric functions
and Hall polynomials}, Second ed., Oxford Univ. Press,
New York - London, 1995.

\bibitem{SS} S.~Saito and N.~Saitoh, {\it Linearization of bilinear
difference equations}, Phys. Lett. {\bf A120} (1987) 322-326; {\it
Gauge and dual symmetries and linearization of Hirota's bilinear
equations}, J. Math. Phys. {\bf 28} (1987) 1052-1055.

\bibitem{Krichever06} I.~Krichever, {\it Characterizing
Jacobians via trisecants of the Kummer Variety}, arXiv.org:
math.AG/0605625.









\bibitem{Shinzawa-1} N.~Shinzawa and S.~Saito,
{\it A symmetric generalization of linear B\"acklund transformation
associated with the Hirota bilinear difference equation}, J. Phys.
A: Math. Gen. {\bf 31} (1998) 4533-4540, arXiv.org:
solv-int/9801002.

\bibitem{Shinzawa-2}
N.~Shinzawa, {\it Symmetric linear B\"acklund transformation for
discrete BKP and DKP equations}, J. Phys. A: Math. Gen. {\bf 33}
(1998) 3957-3970, arXiv.org: solv-int/9907016; N.~Shinzawa and
R.~Hirota, {\it The B\"acklund transformation equations for the
ultradiscrete KP equation}, arXiv.org: nlin.SI/0212014.




\bibitem{KulResh2} P.~Kulish and N.~Reshetikhin,
{\it Diagonalization of $GL(N)$ invariant transfer matrices and
quantum $N$-wave system (Lee model)}, J. Phys. A: Math. Gen. {\bf
16} (1983) L591-L596.


\end{thebibliography}
\end{document}